\documentclass[aps,pre,preprint,amsmath,nofootinbib,showkeys]{revtex4-1}
\usepackage{graphicx}
\usepackage{mathtools} 
\usepackage{lineno}

\begin{document}


\title{Chaotic Lagrangian models for turbulent relative dispersion}


\author{Guglielmo Lacorata}
\email[]{G.Lacorata@isac.cnr.it}
\affiliation{CNR - Istituto di Scienze dell'Atmosfera e del Clima, Via Monteroni, I-73100, Lecce, Italy.}
\author{Angelo Vulpiani}
\email[]{Angelo.Vulpiani@roma1.infn.it}
\affiliation{Dipartimento di Fisica, Universit\'a "La Sapienza", and CNR-ISC, P.le Aldo Moro 2, I-00185 Roma, Italy, and \ 
Kavli Institute for Theoretical Physics, Beijing 100190, China.}

\date{\today}

\begin{abstract} 
\noindent   
A deterministic multi-scale dynamical system is introduced and discussed as prototype model for relative dispersion in stationary, homogeneous and isotropic turbulence.  Unlike stochastic diffusion models,  here trajectory transport and mixing properties are entirely controlled by Lagrangian Chaos.  The anomalous "sweeping effect",  a known drawback common to kinematic simulations, is removed thanks to the use of quasi-Lagrangian coordinates.   Lagrangian dispersion statistics of the model are accurately analyzed by computing the Finite-Scale Lyapunov Exponent (FSLE),  which is the optimal measure of the scaling properties of dispersion.   FSLE scaling exponents provide a severe test to decide whether model simulations  are in agreement with theoretical expectations and/or  observation. The results of  our numerical experiments  cover a wide range of  ``Reynolds numbers''  and show that chaotic deterministic flows can be very efficient, and numerically low-cost, models of turbulent trajectories in stationary, homogeneous and isotropic conditions. The mathematics of the model is relatively simple and, in a geophysical context, potential applications may regard small-scale parameterization issues in general circulation models, mixed layer and/or boundary layer turbulence models  as well as Lagrangian predictability studies.      
 
\end{abstract}


\maketitle 

\section{Introduction}

The physical characteristics of a fluid dynamical system can be classified in two major categories: Eulerian,  dealing with vector and scalar fields as functions of space and time coordinates, and Lagrangian,  dealing with quantities related to the motion of fluid particles. As far as transport processes are concerned, the Lagrangian approach provides information not trivially obtainable from the only knowledge of the velocity field \citep{CFPV:1991}.  The existence of Lagrangian Chaos \citep{Aref:1984, Ottino:1989},  has  shown the possibility to have efficient transport and dispersion even in regular velocity fields, e.g. periodic  in space and/or time.  In addition to this, turbulence \citep{Frisch:1995} can act further  as mechanism of particle dispersion on intermediate scales of motion. 

Modelling Lagrangian particle trajectory evolution is, therefore, an intriguing task both from the theoretical and methodological points of view. 
Ideally, starting from the knowledge of the velocity field, in every point and at every time, one can numerically integrate fluid particle trajectories by means of a computer over arbitrarily long time intervals. 
 Of course, such an approach cannot be used in realistic situations, since models and/or observative data are always affected by finite resolution. 
Trying to accurately describe the dynamics at all scales of motion presents, in general, insurmountable difficulties, even in the hypothesis of having a full knowledge of the physics of the system (which, in actual fact, is only partly true).  
So, usually, one has to find  a suitable compromise between the level of resolution of the model and the number of degrees of freedom to take into account. 

As far as Lagrangian dynamics is concerned, i.e. the evolution of fluid particle trajectories, there are two major statistics to consider: 
one-particle, or absolute, dispersion, i.e. the mean displacement from the initial positions, which depends essentially on the large-scale dynamical features, 
and two-particle, or relative, dispersion, i.e. the mean separation between trajectories, which carries on interesting information about the physics 
of the system \citep{LaCasce:2008}.  In the present paper we will deal only with relative dispersion modelling. 
At this regard, we recall that, in some sense one could say luckily, the Lagrangian properties of a fluid, e.g. mixing and diffusion, do not depend strongly 
on the details of the velocity field, but do depend, indeed,  mainly on the relationship between the characteristic spatial and temporal scales 
of the system \citep{CFPV:1991}.  
As an example, if we consider the diffusion process simulated by a $\tau-$correlated Langevin equation (as in the classical Brownian motion model), such that 
$\sigma^2 \tau \sim D$, where $D$ is the diffusion coefficient, $\sigma^2$ is the variance of the velocity fluctuations and $\tau$ is the velocity auto-correlation time, we notice 
that the same type of dispersion process, i.e. ballistic regime for times smaller than $\tau$ and asymptotic diffusive regime for times larger than $\tau$, 
can be simulated by the chaotic, deterministic scattering across a lattice of (unsteady) kinematic eddies, as long as 
$l^2 / \tau \sim D$, where $l$ and $\tau$ indicate the eddy size and turnover time, respectively. We will come back to this point later.      
 Such a fact opens interesting possibilities from the Lagrangian modelling viewpoint, since there is no need to compute a realistic turbulent 
flow if the scope is {\it only} that to simulate the  the relative dispersion process in turbulent flows. For example, synthetic or kinematic models can accomplish this task with 
good efficiency  and low computational cost, provided some important aspects are suitable considered.     
 The scope of this work is to present and discuss a general kinematic model 
that can be utilized to numerically simulate Lagrangian trajectories in an ideal turbulent environment in which 
stationarity, isotropy and homogeneity conditions are assumed to be fulfilled.      
 What is known as dynamical system approach to Lagrangian transport and mixing \citep{Aref:1984,Ottino:1989,CFPV:1991,BJPV:1998,Ott:2002,WP:2008} will be adopted as guiding line throughout this paper. 

The idea of  implementing kinematic simulations of turbulence, as alternative to stochastic Lagrangian modelling based on Langevin equations \citep{Thomson:1987}, is not new, and many papers  can be counted in support of this strategy \citep{FHMP:1992,FV:1998,NV:2003,OVSH:2006,NN:2011}. 
 Although our kinematic model differs from similar models used in the cited literature,  since it is purely deterministic, the philosophy  of the approach 
is quite the same:   
using an analytical velocity field to numerically compute trajectories that behave, in some statistical sense, like a real turbulent flow. 
Some Authors \citep{TD:2005,DT:2009,EB:2013} have raised objections as far as the use of kinematic simulations is concerned. It was correctly argued, indeed, that a kinematic velocity field, characterized by structures (eddies) that do not drift all over the domain, therefore opposing an unnatural resistance to the  dragging force exerted by the large-scale energetic field, cannot generate trajectories that simulate turbulent diffusion in the correct way. It is known that pair dispersion, within the inertial range of scales where the turbulent cascade has developed, is expected to evolve according to the so-called locality hypothesis, 
i.e. it is assumed that only those structures having characteristic size of the same order as the particle  separation scale contribute efficiently to the dispersion process \citep{Frisch:1995}.  
This means that, e.g., the mean square particle displacement is expected to obey the well known "$t^3$" Richardson-Obukhov law \citep{Richardson:1926}.

In a kinematic model, this property can be obscured (or distorted) by the so-called sweeping effect due to the most energetic components of the velocity field. These large-scale structures drag the particle pairs across the domain but do not have any effect on the kinematic eddies at smaller scales.  This implies that a particle pair does not spend sufficient time around a local eddy such as to ``thermalize'' to the local dynamics. As a consequence,  the mean square relative dispersion deviates from the ``$t^3$'' law of an amount that grows macroscopically in the limit of long inertial range and high time resolution, as explained and discussed in detail by Thomson and Devenish (2005)  \citep{TD:2005}.

In the meanwhile, some years ago, Lacorata et al. \citep{LMR:2008} have first applied a type of   "correct" kinematic model as sub-grid-scale parameterization of two-particle dispersion in a Large-Eddy Simulation of  planetary boundary layer turbulence.  Later, this methodology has been extended also to oceanographic applications \citep{Palatella:2014,LPS:2014}.  The  aim of the present work is to offer a further and specific contribution to clarify this question.  A very general formulation of a three-dimensional, multi-scale, deterministic kinematic model will be introduced and analyzed in detail by means of numerical simulations.    
The Lagrangian turbulence model is built up assuming the validity of K41 theory \citep{Frisch:1995}.  
For many applications, this does not have to be considered as a limitation.    
A severe test on the skills of the model is performed by computing the Finite-Scale Lyapunov Exponents (see next section),  which  measure the spectrum of relative dispersion rates  
at all scales of motion and allow a direct comparison with the theory.   

This paper is organized as follows: 
the FSLE technique is recalled in Section \ref{sec:fsle}; in Section \ref{sec:model}, the Kinematic Lagrangian Model is introduced and discussed; 
the results of the numerical simulations are described   
in Section \ref{sec:results} and a summary of the conclusions that can be drawn at the end of this work is reported in Section \ref{sec:conclusions}.   
  
\section{Relative dispersion: the Finite-Scale Lyapunov Exponent}
\label{sec:fsle}

The Finite-Scale Lyapunov Exponent (FSLE) is a scale-dependent measure of the mean growth 
rate of the distance between two trajectories, an established technique employed in a wide range of applications  \citep{ABCPV:1996,ABCPV:1997,BC:2000,LCO:2003,LALV:2004,LaCasce:2008,BDSLV:2011,LE:2012,ELDN:2014},  from studies on the growth of finite perturbations in chaotic dynamical systems to turbulent pair dispersion in 
ocean and atmosphere, as well as in fluid dynamics laboratory experiments. 
The idea of measuring the diffusivity, i.e. the time derivative of the mean square particle displacement, as function 
of the particle separation actually dates back to the times of Richardson \citep{Richardson:1926}. The FSLE adopts the same philosophy, i.e. it measures the dispersion rate at fixed scale, thus avoiding  issues related to the fixed time averaging procedure \citep{BCCLV:2000}.    
Although definition and properties of the FSLE have already been described and discussed in detail in literature,
 e.g., see \citep{CV:2013} for a very general review,  for the sake of self-consistency we recall here its basic characteristics.     
Given two trajectories, separated at time $t$ by a distance $r(t)$, let us define  $\tau$ as the time interval during which the distance grows from 
$\delta$ to $\varrho \cdot \delta$, where $\varrho \gtrsim 1$. 
Keeping $\delta$ and $\varrho$ fixed, the mean growth time $\langle \tau \rangle$ from $\delta$ to $\varrho \cdot \delta$,  suitably averaged \citep{CV:2013},  defines the FSLE $\lambda(\delta)$ according to the formula: 
\begin{equation} 
\lambda(\delta) \equiv \dfrac{1}{\langle \tau(\delta) \rangle} \, \ln \dfrac{\varrho \cdot \delta}{\delta}.
\label{eq:fsle}
\end{equation}
The quantity $\tau(\delta)$ is the first exit time of the distance $r$ from scale $\delta$ to scale $\varrho \cdot \delta$. It is implicitly assumed that $\langle \tau(\delta) \rangle$, i.e. the phase space average or, equivalently, the average over an arbitrarily large number of numerical experiments, is a function of $\delta$, for $\varrho \sim O(1)$, but it does not depend on $t$, under the hypothesis of stationary statistics.   
The idea is to define $N_s + 1$ scales, such that $\delta_n = \varrho \cdot \delta_{n-1}$, with $n=1,...,N_s$, and to 
compute the FSLE (\ref{eq:fsle}) for each of them. In a Lagrangian dynamical context, the phase space coincides with the familiar physical space and the trajectories are those of passive tracer particles in a given velocity field. 
Normally, the initial scale $\delta_0$ should be smaller than the least characteristic length of the flow, e.g. the Kolmogorov scale, and the final scale $\delta_{N_s}$ should be of the order of the integral scale of the flow or possibly larger, depending on the size of the system.  A common value of the amplification ratio, adopted also in this work, is $\varrho=\sqrt{2}$.  
It is worth stressing a couple of remarks: the time $\tau(\delta)$ is to be computed as the first crossing time of the scale $\varrho \cdot \delta $, starting from the scale $\delta$;  the factor $\varrho$ must be chosen 
not too close to unity, in order to avoid saturation issues of the FSLE at small scale separations due to the finite time resolution of the trajectories\footnote{The algorithm used for the computation of the FSLE is anyway prepared to compensate possible clipping effects \citep{BCCLV:2000}}, and not too larger than unity, otherwise an accurate scale dependent description of the dispersion regimes could not be feasible.

For a fully developed turbulent flow, in which a self-similar energy cascade establishes in the inertial sub-range between a forcing scale $L_0$ (integral length scale) and a dissipative scale $\eta$ (Kolmogorov length scale),  three major regimes are expected to be observable: 

a) exponential separation, $\langle r(t)^2 \rangle \sim r(0)^2 {\rm e}^{2 \lambda_L t}$, on scales smaller than the Kolmogorov length,  $r \ll \eta$, where $\lambda_L$ is the Lagrangian Maximum Lyapunov Exponent\footnote{Taking into account intermittent fluctuations of the effective 
Lyapunov exponent one has $\langle r(t)^2 \rangle \sim r(0)^2 {\rm e}^{L(2) \, t}$, where $L(2) \geq 2\lambda_L$. In the present work we will ignore this detail.} of the flow \citep{CFPV:1991}; 

b) Richardson-Obukhov scaling $\langle r(t)^2 \rangle \sim C_R \, \epsilon \, t^3$ in the inertial range, $\eta < r < L_0$, 
where $\epsilon$ is the rate of energy dissipation of the cascade and $C_R$ is the non-dimensional Richardson's constant 
\citep{Richardson:1926}; henceforth we will refer to any power law of the type $\langle r(t)^2 \rangle \sim t^\nu$, for which $\nu > 1$, as super-diffusion or, equivalently, super-diffusive regime; 
 
c) Standard diffusion $\langle r(t)^2 \rangle \sim 4 \, D_E \, t$ on scales larger than the integral length, $r \gg L_0$, 
where $D_E$ is the eddy diffusion coefficient \citep{Taylor:1921}. 

\begin{figure*}
 \includegraphics[width=25pc,angle=0.]{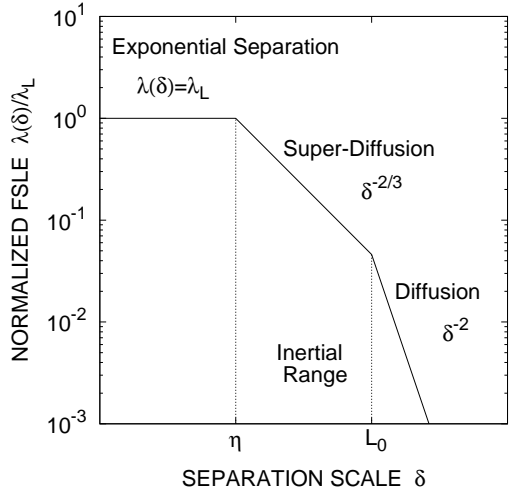}%
\vspace*{2cm}
 \caption{ The three regimes of the normalized FSLE $\lambda(\delta)/\lambda_L$, for an ideal 3D fully developed turbulent flow: a) exponential separation, $\lambda(\delta)=\lambda_L$, for $\delta \ll \eta$, where $\eta$ is the Kolmogorov length; b) Richardson-Obukhov Law, $\lambda(\delta) \sim \delta^{-2/3}$, inside the inertial range $\eta < \delta < L_0$, where $L_0$ is the integral length; c) standard eddy-diffusion, $\lambda(\delta) \sim \delta^{-2}$, at large separation scales $\delta \gg L_0$. 
 } 
 \label{fig:regimes}
 \end{figure*}

Let us briefly discuss each of these regimes and the relationship with the scaling properties of the FSLE, see Figure \ref{fig:regimes}:   

a) the small-scale exponential separation is characterized by a scale-independent mean growth rate of the distance between 
two trajectories. In terms of FSLE this corresponds to $\lambda(\delta)=\lambda_L$ for $\delta \ll \eta$; 

b) by dimensional arguments it can be shown that, inside the inertial range $\eta < \delta < L_0$, the Richardson-Obukhov law corresponds to the scaling $\lambda(\delta) \sim (C_R \cdot \epsilon)^{1/3} \, \delta^{-2/3}$; 

c) as for the previous point, by dimensional arguments it is possible to show that, in the limit of large separation scales 
$\delta \gg L_0$, standard diffusion corresponds to the scaling $\lambda(\delta) \sim D_E \, \delta^{-2}$.  

From the FSLE analysis it is possible not only to distinguish the type of regime a), b) or c) in correspondence of a given range of scales, but also to estimate, at least as order of magnitude, the characteristic parameters of each regime, $\lambda_L$, $\epsilon$ or $D_E$. 
At this regard,  it is clear that an unequivocal measure of the scaling exponents, characteristic of the  relative dispersion process, is needed in order to assess whether or not numerical experiments are in agreement with the theory. 
 We would like to stress again that the FSLE approach, where the "proper" variable is $\delta$, is rather different than looking at $\langle r^2(t) \rangle$ vs $t$. 
In the latter case often, at a given time, one can have contamination of effects due to different scales. As a result of such a contamination,  $\langle r^2(t) \rangle$ can strongly depend on $\langle r^2(0) \rangle$  \citep{CV:2013}. This drawback is ruled out by the FSLE analysis technique.

\section{Kinematic Lagrangian Model}
\label{sec:model}

\subsection{Velocity field}

Let us assume that the flow is incompressible. For a 2D velocity field, 
 ${\bf u}=\{ v_1,v_2 \}$  is a continuous function of the spatial coordinates ${\bf x}=\{ x_1,x_2 \}$ and of the time $t$. 
 The incompressibility condition $\nabla \cdot {\bf u} = 0$ can be fulfilled by  introducing the stream-function $\Psi(x_1,x_2,t)$ and, then,  writing the velocity field as:
\begin{equation}
{\bf u} = (\dfrac{\partial \Psi}{\partial x_2},-\dfrac{\partial \Psi}{\partial x_1})
\label{eq:2D}
\end{equation}
The evolution of the Lagrangian coordinates of a fluid particle is given by:
\begin{equation}
\dfrac{d{\bf r}}{dt} = {\bf u}
\label{eq:2Dlag}
\end{equation}
The dynamical system (\ref{eq:2Dlag}), with the velocity field defined in (\ref{eq:2D}), is formally a Hamiltonian system with $\Psi$ playing the role of the Hamiltonian function \citep{Ottino:1989}.  
Once the stream-function, or equivalently the velocity field, is assigned, the system formed by (\ref{eq:2D}) and (\ref{eq:2Dlag}) is called kinematic model.  
Lagrangian trajectories computed in a 2D kinematic model are regular, i.e. non chaotic, if ${\bf u}$ does not depend explicitly on time $t$. If, conversely, ${\bf u}$ is a nonlinear vector field explicitly depending on the time $t$, Lagrangian dynamics (\ref{eq:2Dlag}) can be chaotic, i.e. two arbitrarily close trajectories tend to separate exponentially in time: 
\begin{equation}
\langle ||\delta {\bf x}(t)|| \rangle \sim ||\delta {\bf x}(0)|| {\rm e}^{\lambda_L t}
\label{eq:mle}
\end{equation}   
in the limit $||\delta {\bf x}|| \to 0$, where $||\delta {\bf x}|| \equiv || {\bf x}^{(1)}-{\bf x}^{(2)} ||$ is the distance between two trajectories ${\bf x}^{(1)}$, ${\bf x}^{(2)}$, and 
$\lambda_L > 0$ is the Maximum Lagrangian Lyapunov Exponent (LLE) of the flow. 
The LLE $\lambda_L$ can be positive even when ${\bf u}$ is a regular function, e.g.  
periodic in space and time. A paradigmatic example of a 2D chaotic flow is the Rayleigh-B\'enard 
convection model by Solomon and Gollub \citep{SG:1988,CFPV:1991}. 

When perturbing an integrable Hamiltonian system, the transition from non chaotic to full chaotic regime occurs via progressive destruction of the KAM tori, until a fully chaotic regime is reached for suitable values of the perturbation parameters, i.e. in the so-called overlap of the resonances regime \citep{Chirikov:1979}. Therefore, typically, chaotic motions co-exist with islands of regular motions in the phase space,  
and the maximum Lyapunov exponent is not positive everywhere but, averaging over all possible initial conditions, it has positive mean \citep{Ottino:1989, Ott:2002}.  In the case discussed in the present work, the model velocity field is suitably perturbed in order to reach the Chirikov regime, without trapping regions or other kind of regular islands.

Let us consider, now, the general three dimensional case.  
Let ${\bf \Psi}=\{\Psi_1,\Psi_2,\Psi_3 \}$  be a continuous and differentiable vector field, function of the spatial coordinates, ${\bf x}=\{ x_1,x_2,x_3 \}$, and of the time, $t$.   
Let ${\bf u}=\{ u_1,u_2,u_3 \}$ be a velocity field defined as ${\bf u}={\bf \nabla} \times {\bf \Psi}$, i.e.: 
\begin{equation}
u_{i_1}  =  \dfrac{\partial \Psi_{i_3}}{\partial x_{i_2}} -   \dfrac{\partial \Psi_{i_2}}{\partial x_{i_3}}
\label{eq:VEL3D} 
\end{equation}
where $\{ i_1,i_2,i_3\}$ stand for all cyclic permutations of $\{1,2,3\}$.   
Of course, ${\bf u}$ is a solenoidal field, i.e. ${\bf \nabla} \cdot {\bf u} \equiv 0$. 
It must be observed that, in the three dimensional case, providing ${\bf u}$ is non linear, a dynamical system of the type (\ref{eq:2Dlag})  
can display Lagrangian chaos even when ${\bf u}$ is not explicitly time dependent \citep{CFPV:1991}. 
Henceforth we will consider only non linear velocity fields that are periodic functions both in space and time. 
At this regard, for simplicity of notation, let us define the following change of variables from ${\bf x}=\{ x_1,x_2,x_3 \}$ to ${\bf s}(t)=\{ s_1(t),s_2(t),s_3(t) \}$ as:
\begin{equation}
s_{i}(t) \equiv  x_i - \varepsilon \, \sin(\omega_i t)  
\label{eq:timepert} 
\end{equation}
where $i=\{1,2,3\}$,  
${\bf x}$ is the position vector in the fixed reference frame, and ${\bf s}(t)$ oscillates periodically around 
${\bf x}$ with frequency ${\bf \omega}/ (2 \pi)$ and amplitude $\varepsilon$. For technical reasons that will be clear later 
it is convenient to define the same amplitude $\varepsilon$ for all components, and the pulsations $\omega_1 \simeq \omega_2 \simeq \omega_3 \simeq \omega$ nearly equal to one another.    
Let us now define a 3D periodic lattice of convective cells as:  
\begin{equation}
\Psi_{i_1}(x_{i_2},x_{i_3},t)  =  \dfrac{A}{k} \sin [ k s_{i_2}(t) ]\sin [ k s_{i_3}(t) ]
\label{eq:PSI3D} 
\end{equation} 
where, as for Eq. (\ref{eq:VEL3D}), the set $\{i_1,i_2,i_3\}$ is a cyclic permutation of $\{1,2,3\}$; 
$k= 2 \pi / l$ and $l$ are spatial wave-number and wavelength, respectively; $\varepsilon$ and $\omega/(2 \pi)$, contained 
in the time dependent terms, $s_1(t)$, $s_2(t)$ and $s_3(t)$, are amplitude and frequency of the stream-function  oscillation. 
The parameter $A$ defines the velocity scale of the flow. From (\ref{eq:VEL3D}) and (\ref{eq:PSI3D}) the three components of the kinematic velocity field, adopting the cyclic index notation as above, turn out to be: 
\begin{widetext}
\begin{equation}
u_{i_1}[{\bf x},t]  = A \{ \sin[k s_{i_1}(t)] \cos[k s_{i_2}(t)] - \sin[k s_{i_1}(t)] \cos[k s_{i_3}(t)] \} 
\label{eq:VEL3Dc} 
\end{equation}
\end{widetext}
The field (\ref{eq:VEL3Dc}) can be seen as a three dimensional version of the Solomon and Gollub model \citep{SG:1988}. 
Other options are possible, of course, in the choice of the analytical form of ${\bf u}$. The ABC flow \citep{CFPV:1991}, for example, or the so-called Double Stream Function (DSF) model \citep{LMR:2008} 
are other good candidates. 
In analogy with the overlap of the resonances in Hamiltonian dynamics \citep{Chirikov:1979}, we must expect 
that, for a certain range of values of the lattice oscillation parameters $\varepsilon$ and $\omega$, every possible bounded  region of regular motion is destroyed and Lagrangian trajectories can evolve chaotically across the whole domain.  
In other words,  for any $\omega$, $\varepsilon$ must be larger than a certain critical threshold $\varepsilon_c(\omega)$  in order to let the model reach  a suitable  {\it working point}, i.e. a regime in which Lagrangian chaos attains  a good efficiency as mechanism of trajectory dispersion. 
At this regard, the standard set up of the parameters can be defined on the basis of the following considerations. 
The 3D periodic lattice is formed by cubic elementary cells of edge $l/2$, and characteristic 
time scale $t_c = 2 \, l \, / \, A$. The periodic oscillations of the structure let a fluid particle escape from 
its initial cell and perform a  sort of random walk through the lattice, with characteristic time of the same order as $t_c$, providing 
the oscillation frequency is of the same order as the turnover frequency, i.e. $\omega/(2 \pi) \sim 1/t_c$. The oscillation 
amplitude can be fixed to some fraction of the characteristic length of the cells, $\varepsilon/l \sim 10^{-1}$. 

In all cellular flows of this type, characterized by only one scale of motion, relative dispersion displays two regimes: small-scale exponential separation, 
$$\langle || \delta {\bf x}(t) || \rangle \sim || \delta {\bf x}(0) || {\rm e}^{\lambda_L t},$$  
in the limit $|| \delta {\bf x} || \ll l$, with $\lambda_L \sim O(1/t_c)$, and large-scale standard diffusion, 
$$\langle || \delta {\bf x}(t) ||^2 \rangle \sim 4 \, D_l \, t,$$   
for $|| \delta{\bf x} || \gg l$ \citep{CFPV:1991}.  
In order to simulate an intermediate regime of turbulent dispersion, between these two asymptotic limits,  it is mandatory to 
include different scales of motions in the velocity field. This can be attained by the superposition of a series 
of $N_m$ self-similar spatial modes, each corresponding to a different wave-number. With the cyclic index notation the resulting multi-scale stream-function is:  
\begin{equation}
\Psi_{i_1}(x_{i_2},x_{i_3},t)  =  \sum\limits_{m=1}^{N_m} \Psi_{i_1}^{(m)}(x_{i_2},x_{i_3},t)
\label{eq:PSI3Dmulti} 
\end{equation}
In (\ref{eq:PSI3Dmulti}) the terms in the right-hand side have the same form as in (\ref{eq:PSI3D}) but now the parameters 
depend on the $m$ mode: $A_m$, $k_m$, $l_m$, $\varepsilon_m$ and $\omega_m$.   
The three components of the kinematic velocity field are now obtained from (\ref{eq:VEL3D}) and (\ref{eq:PSI3Dmulti}). 
At this point, the 3D kinematic model contains $N_m$ spatial modes, each having wavelength $l_m$ and characteristic velocity 
$A_m$. The wavelengths can be related to each other by a recursive rule: $l_m=L_0 / a^{(m-1)}$, for $m=1,...,N_m$, and $a > 1$. 
We let the largest scale $L_0$ correspond to the integral length and the smallest one $l_{N_m} \equiv \eta$ to  the 
Kolmogorov length of the flow. With $L_0$ and $\eta$ fixed, the parameter $a$ determines the density of the modes. 
In order to simulate 3D isotropic homogeneous turbulence, the further step is to assign the Kolmogorov scaling \citep{BJPV:1998} to the velocities of the modes: 
\begin{equation}
A_m^2 \equiv 2 \, C_K \,  \epsilon^{2/3} \, k_m^{-5/3} \, \Delta k_m
\label{eq:kolmogorov}
\end{equation}
where $k_m \equiv 2 \pi / l_m$, $\Delta k_m \equiv k_{m+1}-k_m$, $m=1,...,N_m$; $ \epsilon $ is the mean rate of turbulent dissipation and $C_K$ is a non dimensional adjustable constant (the equivalent Kolmogorov constant). 
Eq. (\ref{eq:kolmogorov}) determines how energy is distributed among the spatial modes in the inertial range.  
Amplitude and frequency of the time dependent oscillating terms are set according to $\varepsilon_m / l_m \sim O(10^{-1})$ and 
$\omega_m t_m / (2 \pi) \sim O(1)$, where the ``local'' time scales are defined as $t_m \equiv 2 l_m / A_m$. The values of $\omega_m$ along the three directions $x_1$, $x_2$ and $x_3$, differ a little from each other of ``irrational'' factors numerically very close to 1. This is to avoid possible (even though unlikely) anomalous trapping of particles inside a box which might affect the chaotic diffusive process.  
At this point we have defined a 3D unsteady lattice of convective cells of various wavelengths ranging from $\eta$ to $L_0$, 
in which the velocities of spatial modes are related to the corresponding characteristic lengths by the Kolmogorov scaling. 
 Integrating Lagrangian trajectories  with this kinematic model and  looking at the two-particle dispersion, we  observe  
that, despite the scaling (\ref{eq:kolmogorov}), the mean square separation, or equivalently the FSLE, departs from  
the expected Richardson-Obukhov scaling inside the inertial range, with a discrepancy increasing with the width of the 
inertial range, $L_0/\eta \gg 1$. As argued by Thomson and Devenish \citep{TD:2005}, this is due to the fact that the large-scale 
velocity field advects a particle pair, having an initially small separation, through the domain without advecting the surrounding small-scale structures, the so-called ``sweeping'' problem. As direct consequence, not only the locality hypothesis inside the inertial range is violated, but also the mean rate of exponential separation, at scales smaller than the Kolmogorov 
length, is sensitive to the integral scale and, ultimately, depends on the width of the inertial range. 

The Maximum Lyapunov Exponent,  on the contrary,  is expected to be    
related to the inverse characteristic time of the smallest coherent structure of the flow \citep{BJPV:1998}, i.e. $\lambda_L \sim 1/t_{\eta}$ where $t_{\eta} \sim  \epsilon^{-1/3} \eta^{2/3}$ is the turnover time at the Kolmogorov scale $\eta$. The locality assumption,  also, as far as the inertial range is concerned, is disregarded since particle pairs cannot ``thermalize'' to the local dynamics spending a sufficient time around eddies of the same size as the separation distance. 
The result is that the LLE depends on the large-scale velocity, with all other parameters fixed, and relative dispersion within 
the inertial range follows an anomalous scaling $\langle \delta {\bf x}(t)^2 \rangle \sim t^{\nu}$ with $9/2 \leq \nu \leq 6$, or, equivalently, in terms of FSLE $\lambda(\delta) \sim \delta^{-\mu}$ with $1/3 \leq \mu \leq 4/9$. The reasons why the scaling 
exponents are expected to vary between these limits have been explained by Thomson and Devenish \citep{TD:2005} and will not be repeated here. We are only concerned in evaluating under what conditions inertial range relative dispersion of the model is in agreement with Richardson's law or not.

\subsection{Quasi-Lagrangian coordinates}

In order to  treat the ``sweeping'' problem  in a proper way, all spatial modes except the largest one can be written as functions of  two particle ({\it relative}), instead of single particle ({\it absolute}), coordinates, i.e. making use of the so-called Quasi-Lagrangian frame technique \citep{BCCV:1999, LMR:2008}. The point is to consider two particles at the same time and advect them together. The trajectories of a whole set of tracer particles can be, in this way, integrated pair by pair.

This strategy allows to fix the problem of having flow structures that do not move together with the fluid particles: computing the kinematic velocity field in the reference frame anchored to the center of mass of two particles simulates the simultaneous advection of  kinematic eddies and particle pair. This advection may generated either by an external large-scale velocity field or, simply, by letting the largest velocity mode of the model be a function of single particle coordinates, as is the case treated in this work.

Let ${\bf x}^{(1)}$ and ${\bf x}^{(2)}$ be two fluid particle positions in the fixed reference frame. The vector position of their center of mass is: 
\begin{equation}
 {\bf x}^{(C)} \equiv \dfrac{{\bf x}^{(1)} + {\bf x}^{(2)}}{2}  
\label{eq:C}
\end{equation}     
The $m$ mode velocity component in ${\bf x}^{(j)}$ ($j=1,2$) can be written as:
\begin{widetext}
\begin{equation}
u_{i_1}^{(m)}[{\bf x}^{(j)},t]  = A_m \{ \sin[k_m s_{i_1}^{(j)}(t)] \cos[k_m s_{i_2}^{(j)}(t)] - 
\sin[k_m s_{i_1}^{(j)}(t)] \cos[k_m s_{i_3}^{(j)}(t)] \} 
 \\
\label{eq:VEL3Dmode} 
\end{equation}
\end{widetext}
where the usual cyclic index notation is adopted, and  
\begin{equation}
{\bf s}^{(j)}(t) \equiv {\bf x}^{(j)}- {\boldsymbol\varepsilon} \, \sin (\omega t)
\label{eq:sj}
\end{equation}
with ${\boldsymbol\varepsilon} \equiv \varepsilon (1,1,1)$.  
Let us consider both particles of a pair, ${\bf x}^{(1,2)}$, and rewrite (\ref{eq:sj}) in the reference frame of their center of mass: 
\begin{equation}
{\bf s}^{(1,2)}_{QL}(t) \equiv [{\bf x}^{(1,2)} - {\bf x}^{(C)}]- {\boldsymbol\varepsilon} \, \sin (\omega t)
\label{eq:QL}
\end{equation}  
where ${\bf x}^{(C)}$ is defined in (\ref{eq:C}). Let us write the full kinematic Lagrangian model as follows: 
\begin{widetext}
\begin{equation}
\begin{cases}
{\bf u}[{\bf x}^{(1,2)},t] = {\bf u}^{(1)}[{\bf s}^{(1,2)}(t)] + \sum\limits_{m=2}^{N_m} {\bf u}^{(m)}[{\bf s}^{(1,2)}_{QL}(t)] \\
\\
\dfrac{d {\bf x}^{(1,2)}}{dt} = {\bf u}[{\bf x}^{(1,2)},t]
\end{cases}
\label{eq:QL-model}
\end{equation}
\end{widetext}
We will call (\ref{eq:QL-model}) QL-model, in which every mode except one ($m=1$) is written as function of the Quasi-Lagrangian 
coordinates relatively to a given particle pair. 
In other terms, the QL-model velocity field depends on the coordinates of both particles of a pair, via their center of mass.  
The kinematic field in (\ref{eq:QL-model}) for ${\bf s}_{QL} \equiv {\bf s}$, i.e. setting ${\bf x}_C \equiv 0$ in (\ref{eq:QL}), will be called E-model, in which every mode is written in terms of single particle  coordinates ${\bf x}$. 
It is worth stressing that the formulation in terms of two particle coordinates does not imply that the QL-model is only a two-particle model \citep{BCCV:1999}, 
since the most energetic mode ($m=1$) in (\ref{eq:QL-model}), function of single particle coordinates, simulates      
a large-scale advection acting on the single trajectories. 
We would like to remark that, in many applications, the role of large-scale advection is normally played, for example, by a general circulation model velocity field, so that there is no need to keep the $m=1$ mode as function of single particle coordinates any longer.  
In the next section, the results obtained from the Lagrangian simulations performed with the QL-model and the E-model will show 
the differences between the two types of configuration, as far as relative dispersion is concerned. 
 
\section{Results}
\label{sec:results}

A standard parameter set up of the 4-decade kinematic model is reported in Tab. \ref{tab:parameters}.  The width of the inertial range, in all analyzed cases,  is determined by the integral length $L_0$, at fixed Kolmogorov length $\eta$.  All other parameters are kept fixed in all numerical simulations. 

\begin{table}[h]%
\begin{tabular}{| l  l   l   l |}
\hline
  & & &   \\
 & Kolmogorov constant & $C_K=1$  & \\
  & turbulent dissipation rate  &   $\epsilon=10^{-3}$  &  \\
  & number of modes &     $N_m = 29$   &  \\
  & mode density  &    $a=\sqrt{2}$   & \\
   & integral length  &    $L_0 = 2^{14}$   & \\
   & wavelengths &   $l_m= 2 \, L_0/a^{m-1}$      ($l_1 = 2 \, L_0$)   &  \\
   & wave-numbers &   $k_m=2 \pi / l_m$  & \\
  & square velocities &   ${A_m}^2 = 2 C_K \epsilon^{2/3} {k_m}^{-5/3} \Delta k_m$      ($\Delta k_m = k_{m+1} - k_m$)  &  \\
  & turnover times &   $t_m = 2 \, l_m / A_m$   & \\
  & oscillation amplitudes &   $\varepsilon_m = 0.2 \, l_m$  &  \\
  & oscillation pulsations &    $\omega_m = 4 \, \pi / t_m$      &    \\
  & Kolmogorov length &    $\eta = l_{N_m}/2  = 1$  & \\
 & integral time &    $T_0 = L_0^{2/3} (2 C_K)^{-1/2} \epsilon^{-1/3}$  &  \\
 & integration time step  &    $\Delta t = 10^{-3} \, t_{N_m}$  &\\
  &   &   &  \\
\hline
\end{tabular}
\caption{Standard set-up for the 4-decade inertial range configuration analyzed in this paper. Parameters are expressed in non dimensional units.}
\label{tab:parameters}
\end{table}%

\begin{figure*}
\centerline{a)}
 \hspace{-5.cm}
\includegraphics[trim=0cm 0cm 0cm 0cm, clip=true, scale=0.4,angle=0.]{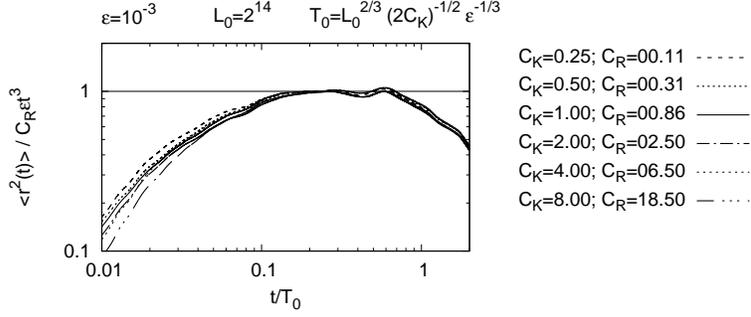}
\centerline{b)} 
\includegraphics[trim=0cm 0cm 0cm 0cm, clip=true, scale=0.4,angle=0.]{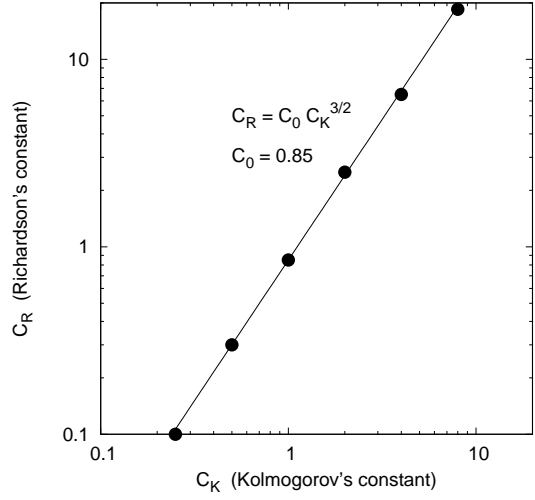}
\caption{Collapse of the rescaled mean square trajectory separations computed in the QL configuration for different values of the Kolmogorov's constant $C_K$.  
Parameters $\epsilon$, $L_0$ and $T_0$ are (expressed in non dimensional units) 
turbulent dissipation rate,  maximum eddy size and integral time scale, respectively (a). Empirical scaling relation between Kolmogorov constant ($C_K$) and Richardson's constant ($C_R$) of the model (b). Richardson's constant $C_R$ is computed by fitting the Richardson's law, at given $C_K$ and $\epsilon$, to the mean square relative separation $\langle r^2(t) \rangle$ inside the inertial range.    
 }  
 \label{fig:cr_ck}
 \end{figure*}

We first checked the properties of relative dispersion in the time domain for the QL-model.  We computed  the mean square pair separation $\langle r(t)^2 \rangle$, for the 4-decade configuration, 
to assess: the existence of the Richardson's scaling law inside the inertial range, at given $\epsilon$; and an empirical relationship between the Richardson's constant 
$C_R$ and the Kolmogorov constant $C_K$ of the model. These results are reported in Fig. \ref{fig:cr_ck}.  We observe that: the QL-model reproduces the expected Richardson's regime for relative dispersion inside the inertial range, at various $C_K$ values; Richardson's constant $C_R$ can be measured and reported as function of $C_K$, at given $\epsilon$; $C_R$ attains values in the expected range,  
$[0.1 - 1]$, for $C_K$ close to the theoretical value, $C_K \simeq 1$. Therefore, in all simulations we fixed $C_K=1$.  We remark also that the empirical scaling relation between the two constants, 
$C_R \sim C_K^{3/2}$, can be justified on the basis of theoretical arguments by noticing that both quantities, $C_R$ and $C_K^{3/2}$, appear as rescaling factor of $\epsilon$ in the Richardson's law and 
in the Kolmogorov spectrum, respectively. 

\begin{figure*}
\centerline{a)}
\hspace*{-3cm}
 \includegraphics[width=22pc,angle=0.]{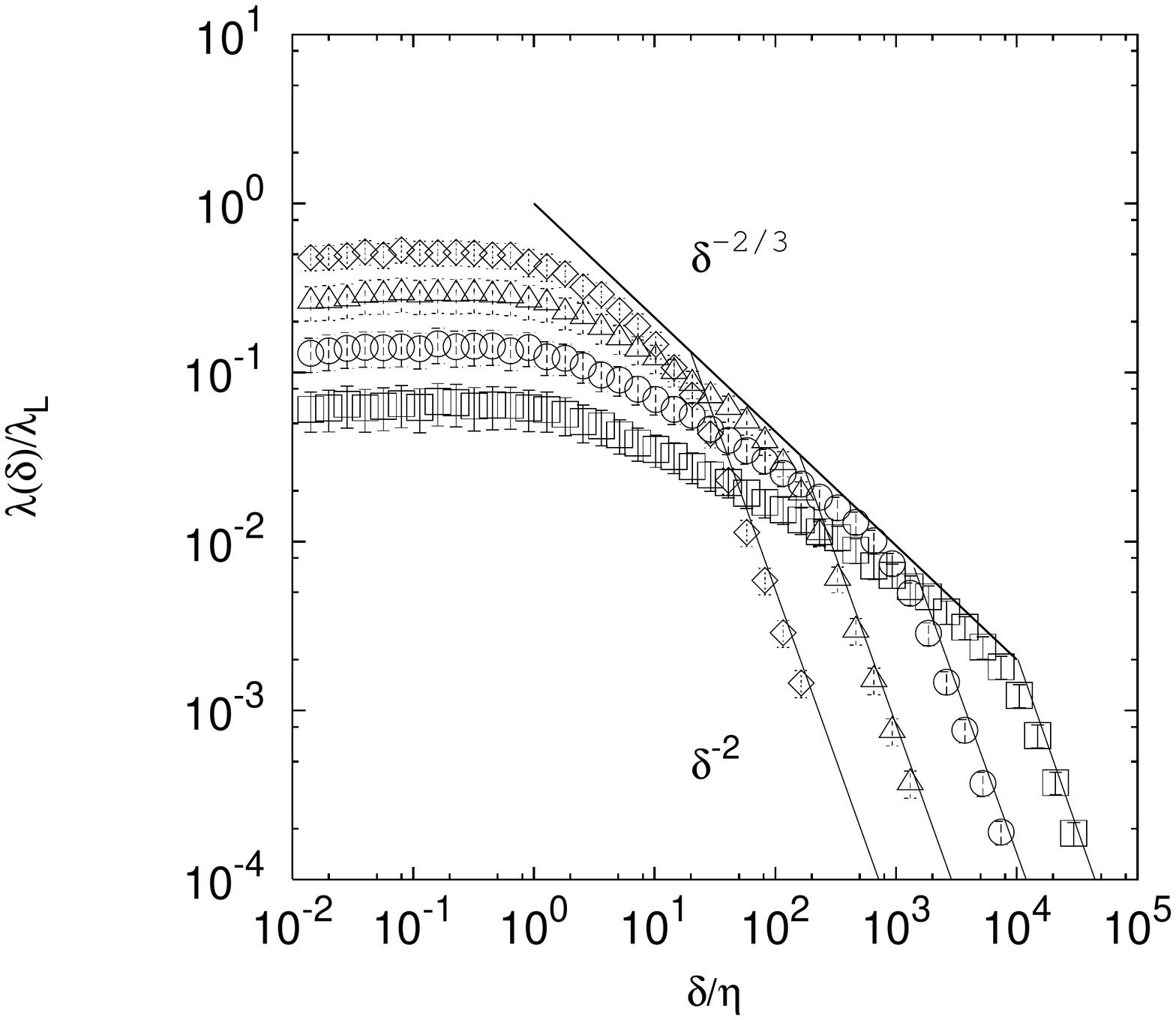}
\centerline{b)}
\hspace*{-3.cm} 
\includegraphics[width=22pc,angle=0.]{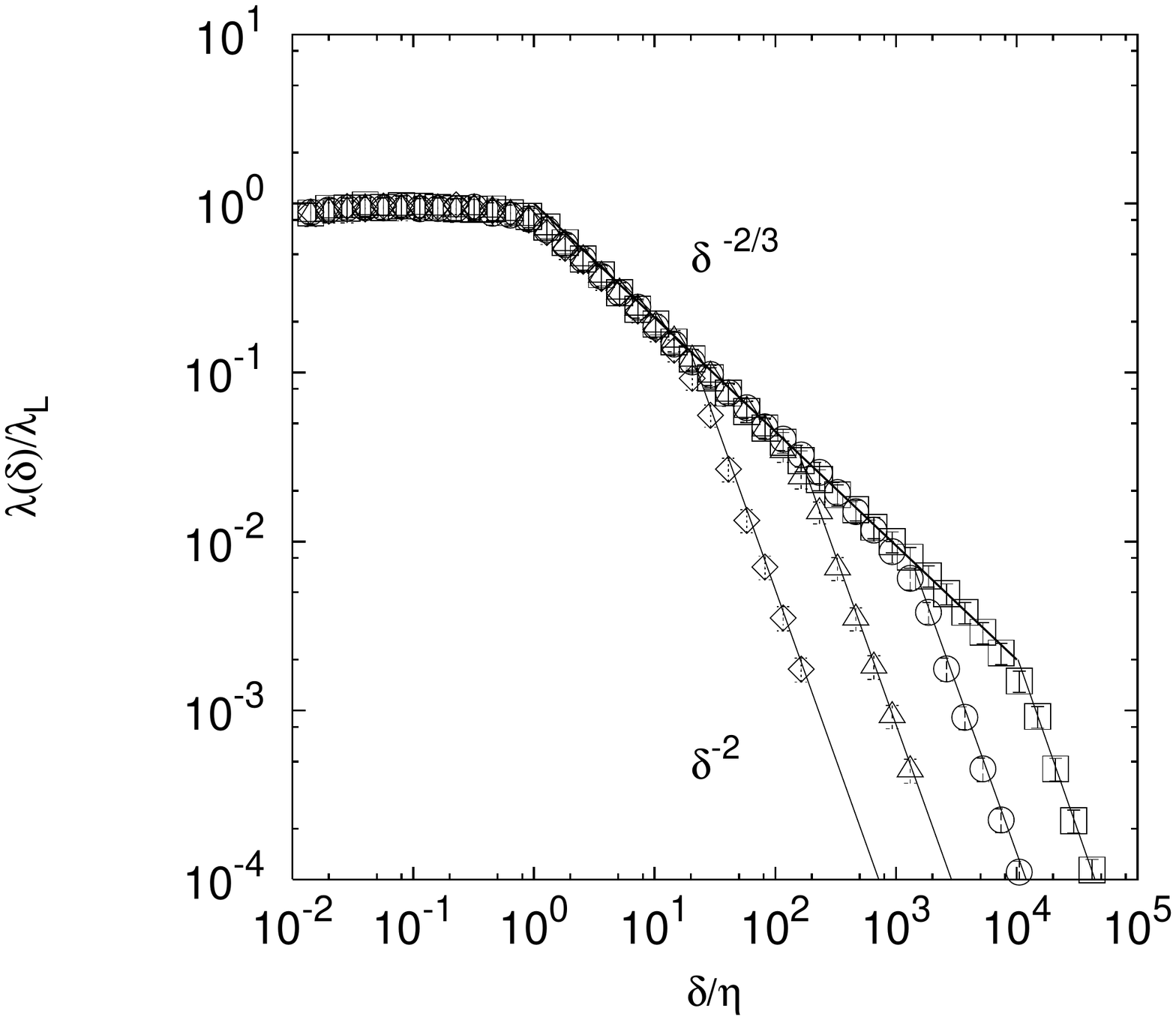}
 \caption{Normalized FSLE $\lambda(\delta)/\lambda_L$, 
  with fixed rate of energy dissipation $\epsilon$,  fixed Kolmogorov length $\eta$ and four different values of the integral length scale $L_0$:  a) $L_0/\eta \sim 10$ (diamonds); b) $L_0/\eta \sim 10^2$ (triangles); c) $L_0/\eta \sim 10^3$ (circles) and d) $L_0/\eta \sim 10^4$ (squares). Statistics over $N=8000$ particle pairs with uniform random initial positions. 
Left panel: E-model; Right panel: QL-model. 
 } 
 \label{fig:4fsle}
 \end{figure*}

FSLE's for E-model and QL-model are plotted in Fig. \ref{fig:4fsle}, for identical parameter set up, varying the width of the inertial range over 
four orders of magnitudes. All curves, normalized to $\lambda_L$, collapse perfectly in the QL-model, fulfilling the invariance property of the Richardson scaling with respect to the inertial range.  On the contrary, in the E-model  such a property does not hold and the scaling deviates progressively from the Richardson's law 
at increasing $L_0/\eta$ ratio. Outside the inertial range,  at scales larger than the integral length, the scaling approaches a standard diffusive regime characterized by the same eddy diffusion coefficient, 
of course $L_0-$dependent, for both models.  At small scales, i.e. below the Kolmogorov length, exponential separation appears but with a mean rate that, in the QL-model, is constant in all numerical experiments while, in the E-model, is sensitive to 
the integral scale. This picture confirms unequivocally that the conjecture advanced by \citet{TD:2005} is verified for what concerns the behavior of kinematic simulations without the  "sweeping effect" correction, and, at the same time, that the Quasi-Lagrangian coordinate technique works perfectly to restore 
the right Richardson's law on arbitrarily long inertial range.


\begin{figure*}
\hspace*{-1.5cm}
 \includegraphics[width=25pc,angle=0.]{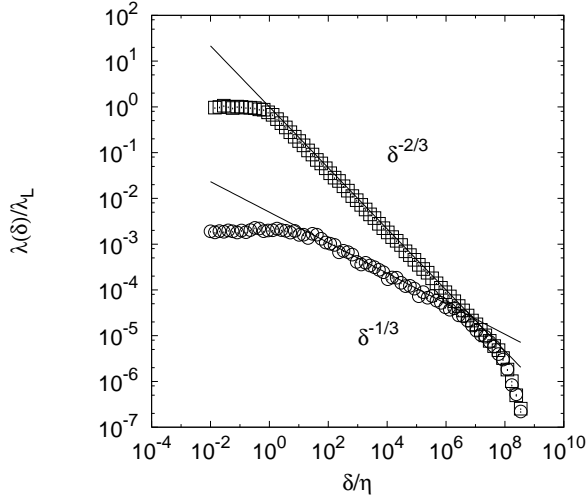}%
 \caption{FSLE for an 8-decade inertial range, $L_0/\eta \sim 10^8$: QL-model (squares) and E-model (circles). 
 Statistical errors are of the same order as the size of the symbols.   
 The QL-model always satisfies the Richardson-Obukhov Law $\sim \delta^{-2/3}$ inside the inertial range while 
 the E-model displays an anomalous $\sim \delta^{-1/3}$.
 } 
 \label{fig:fsle_8dec}
 \end{figure*}

A further  evidence is shown in Fig. \ref{fig:fsle_8dec} where, in order to best highlight the differences, the FSLE's are computed for an eight decade long inertial range in both models. 

The numerical simulations have been performed with both fixed and adaptive integration time step $\Delta t$, showing no difference between the two methods. 
In the fixed time step case, $\Delta t$ is set to a value much  smaller than the Kolmogorov time $t(\eta)$. This 
choice is very conservative since it permits a time resolution of the dispersion process that grows progressively with the scale of motion but, of course, it is also very expensive in terms of computation time when the inertial range attains several decades. On the other hand, the adaptive time step, defined 
as $\Delta t_n=10^{-2} \,  l_n/A_1$ for $l_{n+1} < r < l_n$, is more efficient in terms of resource consuming and give exactly the same results of the former case, regardless the width of the inertial range.  We remark that $l_n/A_1$ is the advecting time at scale $l_n$, and the condition for $dt$ is to be sufficiently small as 
to resolve well the dynamics of dispersion at local scale.  

  \begin{figure*}
\centerline{a)}
\includegraphics[trim=0cm 0cm 0cm -3cm,clip=true,scale=0.3,angle=0.]{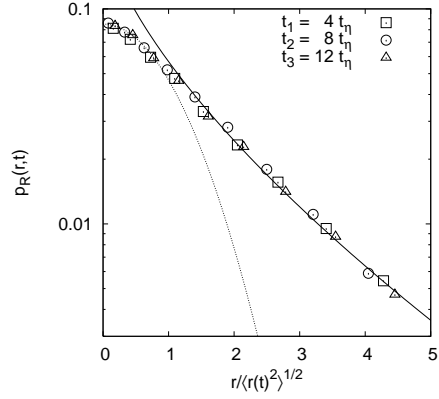}
\centerline{b)}
\includegraphics[trim=0cm 0cm 0cm -1cm,clip=true,scale=0.3,angle=0.]{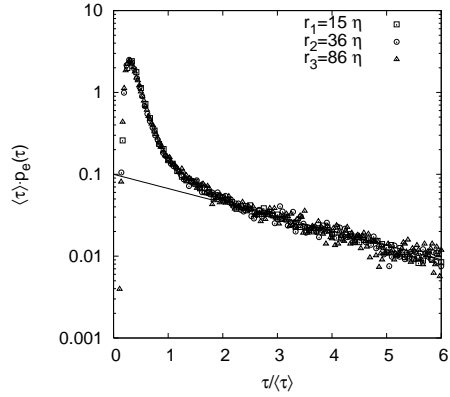}
 \caption{Upper panel:  $p_R(r,t)$ at three different times $t=t_1,t_2,t_3$ measured in Kolmogorov time unit (symbols) as function of $r$ normalized to the standard deviation; the continuous line is the Richardson pdf $p^*_R(r,t)$; the dotted line is a Gaussian pdf.     
 Lower panel: collapse of rescaled $p_e(\tau)$ at three different separations $\delta=r_1,r_2,r_3$ within the inertial range (symbols); the line fitting the exponential tail is 
$p^*_e(\tau)$. 
 } 
 \label{fig:pdfs}
 \end{figure*}

Last, we have  studied the statistics of the dispersion process within the inertial range of the model (with the $L_0/\eta \sim 10^4$ set up, see Fig. \ref{fig:4fsle}) and checked it out with the theoretical predictions. 
At this regard, let us 
define the probability distribution function (pdf) of the distance $r$ between two particles at time $t$ as $p_R(r,t)$, and the pdf of the exit time $\tau$ 
at scale $\delta$ as $p_e(\tau)$.   
Following \citet{BC:2000}, we recall that, in the Richardson regime, 
the expected forms for the two pdf's are, respectively:

\begin{equation}
p^*_R(r,t) \simeq \dfrac{C}{C_R \,  \epsilon  \, t^3} 
{\rm exp}\left( -C'\dfrac{r^{2/3}}{C_R^{1/3} \,  \epsilon^{1/3} \, t}   \right)
\label{eq:richpdf}
\end{equation}

\begin{equation}
p^*_e(\tau) \simeq {\rm exp}\left( -C'' \dfrac{\tau}{\langle \tau(\delta) \rangle} \right) 
\label{eq:fslepdf}
\end{equation}

where $ \epsilon $ is the mean turbulent dissipation rate,  $C_R$ is the Richardson's constant, $\langle \tau(\delta) \rangle$ is the scale-dependent mean exit time,  $C$, $C'$ and $C''$ are constants whose  precise value is not  particularly important in this context.


The results are shown in Fig. \ref{fig:pdfs}. The pdf 
$p_R(r,t)$ is computed at three different times after the release of the particle pairs and, except for the core of the distribution around $r \simeq 0$, the tail is well fitted by the Richardson's  prediction (\ref{eq:richpdf})  up to 4 or 5 times the standard deviation. The pdf $p_e(\tau)$ is computed for three different separation thresholds and, again, except for a limited range of $\tau \simeq 0$, the exponential tail is in agreement with the theoretical prediction (\ref{eq:fslepdf}) as confirmed by the collapse of the re-normalized curves \citep{BC:2000}. 
The discrepancy observed at small $r$ and $\tau$ is likely due to the "smoothness" of the kinematic model at all scale separations with respect to the "roughness" of a real turbulent field.

\section{Discussion and conclusions}
\label{sec:conclusions}

In this paper, the problem of modelling two-particle dispersion in a turbulent fluid has been addressed by means of a dynamical system approach. 
 Unlike other types of kinematic simulations, Lagrangian chaos and deterministic dynamics are the key elements characterizing the model we have here introduced and analyzed.   
The main conclusions at the end of this work can be summarized as follows:  

1) Multi-scale dynamics typical of turbulent flows is guaranteed by the superposition of a number of self-similar modes. 

2) Kolmogorov relationship between space and velocity parameters, scale by scale, define an inertial range in the model. 

3) Quasi-Lagrangian coordinates assure the statistics of turbulent relative dispersion turn out to be in very good agreement with the theoretical expectations.  

While points 1) and 2) are already established facts in literature, point 3) is an original element, fundamental to the correct behavior of the kinematic model.   

We would like to remark also some other points:  

a) The model is conceived to describe two-particle Lagrangian dispersion. This does not have to be considered as a limitation since it can be verified (not shown) that also one-particle, or absolute dispersion, statistics is reproduced coherently with the theory of diffusion. 

b) The statistical quantities characterizing the relative dispersion process are not altered if one considers the separation between particles that belong 
to different pairs. This assures that the properties of dispersion are common to a whole set of tracer particles and are not limited to the single pairs separately. 

c) The kinematic eddies are long-living coherent structures, i.e. with  slowly decaying Eulerian auto-correlations. It can be verified that letting the velocity amplitude 
of the modes evolve in time as a stochastic variable, e.g. via a time-correlated Langevin equation, does not change the Lagrangian characteristics of the dispersion process.  

d) The kinematic model can be used either as independent model of an ideal turbulent flow or as sub-grid or, more in general, small-scale model for 
turbulent dispersion inside some more complex and realistic large-scale model of, e.g., atmospheric or oceanic circulation. At this regard, a two-dimensional version of the model, see \citet{LPS:2014}, can be adapted to improve the simulation of horizontal mesoscale dispersion over large domains. 

e) The scaling properties of the model are not limited to the Richardson's regime but, in principle, any arbitrary scaling law can be imported in the model, depending on the available experimental information about a given system. For modelling dispersion in the ocean upper layer, for example, the 2D version of the model can be set up according to the information coming from Lagrangian drifter data.

\begin{acknowledgments}
GL would like to thank ``RITMARE" (Ricerca ITaliana per il MARE) Reseach Program for providing financial support and scientific motivations for this work. 
 AV thanks the Kavli Institute for Theoretical Physics, CAS, Beijing, for the kind hospitality at the Workshop "Non-equilibrium processes at the nanoscales" (July-August, 2016) where this manuscript was completed.      
\end{acknowledgments}

\newpage

\begin{thebibliography}{35}%
\makeatletter
\providecommand \@ifxundefined [1]{%
 \@ifx{#1\undefined}
}%
\providecommand \@ifnum [1]{%
 \ifnum #1\expandafter \@firstoftwo
 \else \expandafter \@secondoftwo
 \fi
}%
\providecommand \@ifx [1]{%
 \ifx #1\expandafter \@firstoftwo
 \else \expandafter \@secondoftwo
 \fi
}%
\providecommand \natexlab [1]{#1}%
\providecommand \enquote  [1]{``#1''}%
\providecommand \bibnamefont  [1]{#1}%
\providecommand \bibfnamefont [1]{#1}%
\providecommand \citenamefont [1]{#1}%
\providecommand \href@noop [0]{\@secondoftwo}%
\providecommand \href [0]{\begingroup \@sanitize@url \@href}%
\providecommand \@href[1]{\@@startlink{#1}\@@href}%
\providecommand \@@href[1]{\endgroup#1\@@endlink}%
\providecommand \@sanitize@url [0]{\catcode `\\12\catcode `\$12\catcode
  `\&12\catcode `\#12\catcode `\^12\catcode `\_12\catcode `\%12\relax}%
\providecommand \@@startlink[1]{}%
\providecommand \@@endlink[0]{}%
\providecommand \url  [0]{\begingroup\@sanitize@url \@url }%
\providecommand \@url [1]{\endgroup\@href {#1}{\urlprefix }}%
\providecommand \urlprefix  [0]{URL }%
\providecommand \Eprint [0]{\href }%
\providecommand \doibase [0]{http://dx.doi.org/}%
\providecommand \selectlanguage [0]{\@gobble}%
\providecommand \bibinfo  [0]{\@secondoftwo}%
\providecommand \bibfield  [0]{\@secondoftwo}%
\providecommand \translation [1]{[#1]}%
\providecommand \BibitemOpen [0]{}%
\providecommand \bibitemStop [0]{}%
\providecommand \bibitemNoStop [0]{.\EOS\space}%
\providecommand \EOS [0]{\spacefactor3000\relax}%
\providecommand \BibitemShut  [1]{\csname bibitem#1\endcsname}%
\let\auto@bib@innerbib\@empty
\bibitem [{\citenamefont {Crisanti}\ \emph {et~al.}(1991)\citenamefont
  {Crisanti}, \citenamefont {Falcioni}, \citenamefont {Paladin},\ and\
  \citenamefont {Vulpiani}}]{CFPV:1991}%
  \BibitemOpen
  \bibfield  {author} {\bibinfo {author} {\bibfnamefont {A.}~\bibnamefont
  {Crisanti}}, \bibinfo {author} {\bibfnamefont {M.}~\bibnamefont {Falcioni}},
  \bibinfo {author} {\bibfnamefont {G.}~\bibnamefont {Paladin}}, \ and\
  \bibinfo {author} {\bibfnamefont {A.}~\bibnamefont {Vulpiani}},\ }\href@noop
  {} {\bibfield  {journal} {\bibinfo  {journal} {Riv. Nuovo Cim.}\ }\textbf
  {\bibinfo {volume} {14}},\ \bibinfo {pages} {1} (\bibinfo {year}
  {1991})}\BibitemShut {NoStop}%
\bibitem [{\citenamefont {Aref}(1984)}]{Aref:1984}%
  \BibitemOpen
  \bibfield  {author} {\bibinfo {author} {\bibfnamefont {H.}~\bibnamefont
  {Aref}},\ }\href@noop {} {\bibfield  {journal} {\bibinfo  {journal} {J. of
  Fluid Mech.}\ }\textbf {\bibinfo {volume} {143}},\ \bibinfo {pages} {1}
  (\bibinfo {year} {1984})}\BibitemShut {NoStop}%
\bibitem [{\citenamefont {Ottino}(1989)}]{Ottino:1989}%
  \BibitemOpen
  \bibfield  {author} {\bibinfo {author} {\bibfnamefont {J.~M.}\ \bibnamefont
  {Ottino}},\ }\href@noop {} {\emph {\bibinfo {title} {The kinematics of
  mixing: stretching, chaos and transport}}}\ (\bibinfo  {publisher} {Cambridge
  University Press},\ \bibinfo {year} {1989})\ p.\ \bibinfo {pages}
  {378}\BibitemShut {NoStop}%
\bibitem [{\citenamefont {Frisch}(1995)}]{Frisch:1995}%
  \BibitemOpen
  \bibfield  {author} {\bibinfo {author} {\bibfnamefont {U.}~\bibnamefont
  {Frisch}},\ }\href@noop {} {\emph {\bibinfo {title} {Turbulence, the legacy
  of {A}.{N}. {K}olmogorov}}}\ (\bibinfo  {publisher} {Cambridge University
  Press},\ \bibinfo {year} {1995})\ p.\ \bibinfo {pages} {296}\BibitemShut
  {NoStop}%
\bibitem [{\citenamefont {LaCasce}(2008)}]{LaCasce:2008}%
  \BibitemOpen
  \bibfield  {author} {\bibinfo {author} {\bibfnamefont {J.~H.}\ \bibnamefont
  {LaCasce}},\ }\href@noop {} {\bibfield  {journal} {\bibinfo  {journal}
  {Progress in Oceanography}\ }\textbf {\bibinfo {volume} {77}},\ \bibinfo
  {pages} {1} (\bibinfo {year} {2008})}\BibitemShut {NoStop}%
\bibitem [{\citenamefont {Bohr}\ \emph {et~al.}(1998)\citenamefont {Bohr},
  \citenamefont {Jensen}, \citenamefont {Paladin},\ and\ \citenamefont
  {Vulpiani}}]{BJPV:1998}%
  \BibitemOpen
  \bibfield  {author} {\bibinfo {author} {\bibfnamefont {T.}~\bibnamefont
  {Bohr}}, \bibinfo {author} {\bibfnamefont {M.~H.}\ \bibnamefont {Jensen}},
  \bibinfo {author} {\bibfnamefont {G.}~\bibnamefont {Paladin}}, \ and\
  \bibinfo {author} {\bibfnamefont {A.}~\bibnamefont {Vulpiani}},\ }\href@noop
  {} {\emph {\bibinfo {title} {Dynamical systems approach to Turbulence}}}\
  (\bibinfo  {publisher} {Cambridge University Press},\ \bibinfo {year}
  {1998})\ p.\ \bibinfo {pages} {350}\BibitemShut {NoStop}%
\bibitem [{\citenamefont {Ott}(2002)}]{Ott:2002}%
  \BibitemOpen
  \bibfield  {author} {\bibinfo {author} {\bibfnamefont {E.}~\bibnamefont
  {Ott}},\ }\href@noop {} {\emph {\bibinfo {title} {Chaos in dynamical
  systems}}}\ (\bibinfo  {publisher} {Cambridge University Press},\ \bibinfo
  {year} {2002})\ p.\ \bibinfo {pages} {490}\BibitemShut {NoStop}%
\bibitem [{\citenamefont {Weiss}\ and\ \citenamefont
  {Provenzale}(2008)}]{WP:2008}%
  \BibitemOpen
  \bibinfo {editor} {\bibfnamefont {J.~B.}\ \bibnamefont {Weiss}}\ and\
  \bibinfo {editor} {\bibfnamefont {A.}~\bibnamefont {Provenzale}},\ eds.,\
  \href@noop {} {\emph {\bibinfo {title} {Transport and Mixing in Geophysical
  flows}}},\ Vol.\ \bibinfo {volume} {744}\ (\bibinfo  {publisher} {Springer
  Berlin Heidelberg},\ \bibinfo {year} {2008})\ p.\ \bibinfo {pages}
  {261}\BibitemShut {NoStop}%
\bibitem [{\citenamefont {Thomson}(1987)}]{Thomson:1987}%
  \BibitemOpen
  \bibfield  {author} {\bibinfo {author} {\bibfnamefont {D.~J.}\ \bibnamefont
  {Thomson}},\ }\href@noop {} {\bibfield  {journal} {\bibinfo  {journal}
  {Journal of Fluid Mechanics}\ }\textbf {\bibinfo {volume} {180}},\ \bibinfo
  {pages} {529} (\bibinfo {year} {1987})}\BibitemShut {NoStop}%
\bibitem [{\citenamefont {Fung}\ \emph {et~al.}(1992)\citenamefont {Fung},
  \citenamefont {Hunt}, \citenamefont {Malik},\ and\ \citenamefont
  {Perkins}}]{FHMP:1992}%
  \BibitemOpen
  \bibfield  {author} {\bibinfo {author} {\bibfnamefont {J.~C.~H.}\
  \bibnamefont {Fung}}, \bibinfo {author} {\bibfnamefont {J.~C.~R.}\
  \bibnamefont {Hunt}}, \bibinfo {author} {\bibfnamefont {N.~A.}\ \bibnamefont
  {Malik}}, \ and\ \bibinfo {author} {\bibfnamefont {R.~J.}\ \bibnamefont
  {Perkins}},\ }\href@noop {} {\bibfield  {journal} {\bibinfo  {journal} {J.
  Fluid Mech.}\ }\textbf {\bibinfo {volume} {236}},\ \bibinfo {pages} {281}
  (\bibinfo {year} {1992})}\BibitemShut {NoStop}%
\bibitem [{\citenamefont {Fung}\ and\ \citenamefont
  {Vassilicos}(1998)}]{FV:1998}%
  \BibitemOpen
  \bibfield  {author} {\bibinfo {author} {\bibfnamefont {J.~C.~H.}\
  \bibnamefont {Fung}}\ and\ \bibinfo {author} {\bibfnamefont {J.~C.}\
  \bibnamefont {Vassilicos}},\ }\href@noop {} {\bibfield  {journal} {\bibinfo
  {journal} {Phys. Rev. E}\ }\textbf {\bibinfo {volume} {57}},\ \bibinfo
  {pages} {1677} (\bibinfo {year} {1998})}\BibitemShut {NoStop}%
\bibitem [{\citenamefont {Nicolleau}\ and\ \citenamefont
  {Vassilicos}(2003)}]{NV:2003}%
  \BibitemOpen
  \bibfield  {author} {\bibinfo {author} {\bibfnamefont {F.~C. G.~A.}\
  \bibnamefont {Nicolleau}}\ and\ \bibinfo {author} {\bibfnamefont {J.~C.}\
  \bibnamefont {Vassilicos}},\ }\href@noop {} {\bibfield  {journal} {\bibinfo
  {journal} {Phys. Rev. Lett.}\ }\textbf {\bibinfo {volume} {90}},\ \bibinfo
  {pages} {024503 1} (\bibinfo {year} {2003})}\BibitemShut {NoStop}%
\bibitem [{\citenamefont {Osborne}\ \emph {et~al.}(2006)\citenamefont
  {Osborne}, \citenamefont {Vassilicos}, \citenamefont {Sung},\ and\
  \citenamefont {Haigh}}]{OVSH:2006}%
  \BibitemOpen
  \bibfield  {author} {\bibinfo {author} {\bibfnamefont {D.~R.}\ \bibnamefont
  {Osborne}}, \bibinfo {author} {\bibfnamefont {J.~C.}\ \bibnamefont
  {Vassilicos}}, \bibinfo {author} {\bibfnamefont {K.}~\bibnamefont {Sung}}, \
  and\ \bibinfo {author} {\bibfnamefont {J.~D.}\ \bibnamefont {Haigh}},\
  }\href@noop {} {\bibfield  {journal} {\bibinfo  {journal} {Phys. Rev. E}\
  }\textbf {\bibinfo {volume} {74}},\ \bibinfo {pages} {036309} (\bibinfo
  {year} {2006})}\BibitemShut {NoStop}%
\bibitem [{\citenamefont {Nicolleau}\ and\ \citenamefont
  {Nowakowski}(2011)}]{NN:2011}%
  \BibitemOpen
  \bibfield  {author} {\bibinfo {author} {\bibfnamefont {F.~C. G.~A.}\
  \bibnamefont {Nicolleau}}\ and\ \bibinfo {author} {\bibfnamefont {A.~F.}\
  \bibnamefont {Nowakowski}},\ }\href@noop {} {\bibfield  {journal} {\bibinfo
  {journal} {Phys. Rev. E}\ }\textbf {\bibinfo {volume} {83}},\ \bibinfo
  {pages} {056317 1} (\bibinfo {year} {2011})}\BibitemShut {NoStop}%
\bibitem [{\citenamefont {Thomson}\ and\ \citenamefont
  {Devenish}(2005)}]{TD:2005}%
  \BibitemOpen
  \bibfield  {author} {\bibinfo {author} {\bibfnamefont {D.~J.}\ \bibnamefont
  {Thomson}}\ and\ \bibinfo {author} {\bibfnamefont {B.~J.}\ \bibnamefont
  {Devenish}},\ }\href@noop {} {\bibfield  {journal} {\bibinfo  {journal}
  {Journal of Fluid Mechanics}\ }\textbf {\bibinfo {volume} {526}},\ \bibinfo
  {pages} {277} (\bibinfo {year} {2005})}\BibitemShut {NoStop}%
\bibitem [{\citenamefont {Devenish}\ and\ \citenamefont
  {Thomson}(2009)}]{DT:2009}%
  \BibitemOpen
  \bibfield  {author} {\bibinfo {author} {\bibfnamefont {B.~J.}\ \bibnamefont
  {Devenish}}\ and\ \bibinfo {author} {\bibfnamefont {D.~J.}\ \bibnamefont
  {Thomson}},\ }\href@noop {} {\bibfield  {journal} {\bibinfo  {journal} {Phys.
  Rev. E}\ }\textbf {\bibinfo {volume} {80}},\ \bibinfo {pages} {048301}
  (\bibinfo {year} {2009})}\BibitemShut {NoStop}%
\bibitem [{\citenamefont {Eyink}\ and\ \citenamefont
  {Benveniste}(2013)}]{EB:2013}%
  \BibitemOpen
  \bibfield  {author} {\bibinfo {author} {\bibfnamefont {G.~L.}\ \bibnamefont
  {Eyink}}\ and\ \bibinfo {author} {\bibfnamefont {D.}~\bibnamefont
  {Benveniste}},\ }\href@noop {} {\bibfield  {journal} {\bibinfo  {journal}
  {Phys. Rev. E}\ }\textbf {\bibinfo {volume} {87}},\ \bibinfo {pages} {023011}
  (\bibinfo {year} {2013})}\BibitemShut {NoStop}%
\bibitem [{\citenamefont {Richardson}(1926)}]{Richardson:1926}%
  \BibitemOpen
  \bibfield  {author} {\bibinfo {author} {\bibfnamefont {L.~F.}\ \bibnamefont
  {Richardson}},\ }\href@noop {} {\bibfield  {journal} {\bibinfo  {journal}
  {Proceedings of the Royal Society A}\ }\textbf {\bibinfo {volume} {110}},\
  \bibinfo {pages} {709} (\bibinfo {year} {1926})}\BibitemShut {NoStop}%
\bibitem [{\citenamefont {Lacorata}\ \emph {et~al.}(2008)\citenamefont
  {Lacorata}, \citenamefont {Mazzino},\ and\ \citenamefont {Rizza}}]{LMR:2008}%
  \BibitemOpen
  \bibfield  {author} {\bibinfo {author} {\bibfnamefont {G.}~\bibnamefont
  {Lacorata}}, \bibinfo {author} {\bibfnamefont {A.}~\bibnamefont {Mazzino}}, \
  and\ \bibinfo {author} {\bibfnamefont {U.}~\bibnamefont {Rizza}},\
  }\href@noop {} {\bibfield  {journal} {\bibinfo  {journal} {Journal of the
  Atmospheric Sciences}\ }\textbf {\bibinfo {volume} {65}},\ \bibinfo {pages}
  {2389} (\bibinfo {year} {2008})}\BibitemShut {NoStop}%
\bibitem [{\citenamefont {Palatella}\ \emph {et~al.}(2014)\citenamefont
  {Palatella}, \citenamefont {Bignami}, \citenamefont {Falcini}, \citenamefont
  {Lacorata}, \citenamefont {Lanotte},\ and\ \citenamefont
  {Santoleri}}]{Palatella:2014}%
  \BibitemOpen
  \bibfield  {author} {\bibinfo {author} {\bibfnamefont {L.}~\bibnamefont
  {Palatella}}, \bibinfo {author} {\bibfnamefont {F.}~\bibnamefont {Bignami}},
  \bibinfo {author} {\bibfnamefont {F.}~\bibnamefont {Falcini}}, \bibinfo
  {author} {\bibfnamefont {G.}~\bibnamefont {Lacorata}}, \bibinfo {author}
  {\bibfnamefont {A.~S.}\ \bibnamefont {Lanotte}}, \ and\ \bibinfo {author}
  {\bibfnamefont {R.}~\bibnamefont {Santoleri}},\ }\href@noop {} {\bibfield
  {journal} {\bibinfo  {journal} {Journal of Geophysical Research Ocean}\
  }\textbf {\bibinfo {volume} {119}},\ \bibinfo {pages} {1306} (\bibinfo {year}
  {2014})}\BibitemShut {NoStop}%
\bibitem [{\citenamefont {Lacorata}\ \emph {et~al.}(2014)\citenamefont
  {Lacorata}, \citenamefont {Palatella},\ and\ \citenamefont
  {Santoleri}}]{LPS:2014}%
  \BibitemOpen
  \bibfield  {author} {\bibinfo {author} {\bibfnamefont {G.}~\bibnamefont
  {Lacorata}}, \bibinfo {author} {\bibfnamefont {L.}~\bibnamefont {Palatella}},
  \ and\ \bibinfo {author} {\bibfnamefont {R.}~\bibnamefont {Santoleri}},\
  }\href@noop {} {\bibfield  {journal} {\bibinfo  {journal} {Journal of
  Geophysical Research Ocean}\ }\textbf {\bibinfo {volume} {119}},\ \bibinfo
  {pages} {8029} (\bibinfo {year} {2014})}\BibitemShut {NoStop}%
\bibitem [{\citenamefont {Aurell}\ \emph {et~al.}(1996)\citenamefont {Aurell},
  \citenamefont {Boffetta}, \citenamefont {Crisanti}, \citenamefont {Paladin},\
  and\ \citenamefont {Vulpiani}}]{ABCPV:1996}%
  \BibitemOpen
  \bibfield  {author} {\bibinfo {author} {\bibfnamefont {E.}~\bibnamefont
  {Aurell}}, \bibinfo {author} {\bibfnamefont {G.}~\bibnamefont {Boffetta}},
  \bibinfo {author} {\bibfnamefont {A.}~\bibnamefont {Crisanti}}, \bibinfo
  {author} {\bibfnamefont {G.}~\bibnamefont {Paladin}}, \ and\ \bibinfo
  {author} {\bibfnamefont {A.}~\bibnamefont {Vulpiani}},\ }\href@noop {}
  {\bibfield  {journal} {\bibinfo  {journal} {Phys. Rev. Lett.}\ }\textbf
  {\bibinfo {volume} {77}},\ \bibinfo {pages} {1262} (\bibinfo {year}
  {1996})}\BibitemShut {NoStop}%
\bibitem [{\citenamefont {Aurell}\ \emph {et~al.}(1997)\citenamefont {Aurell},
  \citenamefont {Boffetta}, \citenamefont {Crisanti}, \citenamefont {Paladin},\
  and\ \citenamefont {Vulpiani}}]{ABCPV:1997}%
  \BibitemOpen
  \bibfield  {author} {\bibinfo {author} {\bibfnamefont {E.}~\bibnamefont
  {Aurell}}, \bibinfo {author} {\bibfnamefont {G.}~\bibnamefont {Boffetta}},
  \bibinfo {author} {\bibfnamefont {A.}~\bibnamefont {Crisanti}}, \bibinfo
  {author} {\bibfnamefont {G.}~\bibnamefont {Paladin}}, \ and\ \bibinfo
  {author} {\bibfnamefont {A.}~\bibnamefont {Vulpiani}},\ }\href@noop {}
  {\bibfield  {journal} {\bibinfo  {journal} {Journal of Physics A}\ }\textbf
  {\bibinfo {volume} {30}},\ \bibinfo {pages} {1} (\bibinfo {year}
  {1997})}\BibitemShut {NoStop}%
\bibitem [{\citenamefont {Boffetta}\ and\ \citenamefont
  {Celani}(2000)}]{BC:2000}%
  \BibitemOpen
  \bibfield  {author} {\bibinfo {author} {\bibfnamefont {G.}~\bibnamefont
  {Boffetta}}\ and\ \bibinfo {author} {\bibfnamefont {A.}~\bibnamefont
  {Celani}},\ }\href@noop {} {\bibfield  {journal} {\bibinfo  {journal}
  {Physica A}\ }\textbf {\bibinfo {volume} {280}},\ \bibinfo {pages} {1}
  (\bibinfo {year} {2000})}\BibitemShut {NoStop}%
\bibitem [{\citenamefont {LaCasce}\ and\ \citenamefont
  {Ohlmann}(2003)}]{LCO:2003}%
  \BibitemOpen
  \bibfield  {author} {\bibinfo {author} {\bibfnamefont {J.~H.}\ \bibnamefont
  {LaCasce}}\ and\ \bibinfo {author} {\bibfnamefont {C.}~\bibnamefont
  {Ohlmann}},\ }\href@noop {} {\bibfield  {journal} {\bibinfo  {journal}
  {Journal of Marine Research}\ }\textbf {\bibinfo {volume} {61}},\ \bibinfo
  {pages} {285} (\bibinfo {year} {2003})}\BibitemShut {NoStop}%
\bibitem [{\citenamefont {Lacorata}\ \emph {et~al.}(2004)\citenamefont
  {Lacorata}, \citenamefont {Aurell}, \citenamefont {Legras},\ and\
  \citenamefont {Vulpiani}}]{LALV:2004}%
  \BibitemOpen
  \bibfield  {author} {\bibinfo {author} {\bibfnamefont {G.}~\bibnamefont
  {Lacorata}}, \bibinfo {author} {\bibfnamefont {E.}~\bibnamefont {Aurell}},
  \bibinfo {author} {\bibfnamefont {B.}~\bibnamefont {Legras}}, \ and\ \bibinfo
  {author} {\bibfnamefont {A.}~\bibnamefont {Vulpiani}},\ }\href@noop {}
  {\bibfield  {journal} {\bibinfo  {journal} {Journal of the Atmospheric
  Sciences}\ }\textbf {\bibinfo {volume} {61}},\ \bibinfo {pages} {2936}
  (\bibinfo {year} {2004})}\BibitemShut {NoStop}%
\bibitem [{\citenamefont {Berti}\ \emph {et~al.}(2011)\citenamefont {Berti},
  \citenamefont {{Dos Santos}}, \citenamefont {Lacorata},\ and\ \citenamefont
  {Vulpiani}}]{BDSLV:2011}%
  \BibitemOpen
  \bibfield  {author} {\bibinfo {author} {\bibfnamefont {S.}~\bibnamefont
  {Berti}}, \bibinfo {author} {\bibfnamefont {F.}~\bibnamefont {{Dos Santos}}},
  \bibinfo {author} {\bibfnamefont {G.}~\bibnamefont {Lacorata}}, \ and\
  \bibinfo {author} {\bibfnamefont {A.}~\bibnamefont {Vulpiani}},\ }\href@noop
  {} {\bibfield  {journal} {\bibinfo  {journal} {Journal of Physical
  Oceanography}\ }\textbf {\bibinfo {volume} {41}},\ \bibinfo {pages} {1659}
  (\bibinfo {year} {2011})}\BibitemShut {NoStop}%
\bibitem [{\citenamefont {Lacorata}\ and\ \citenamefont
  {Espa}(2012)}]{LE:2012}%
  \BibitemOpen
  \bibfield  {author} {\bibinfo {author} {\bibfnamefont {G.}~\bibnamefont
  {Lacorata}}\ and\ \bibinfo {author} {\bibfnamefont {S.}~\bibnamefont
  {Espa}},\ }\href {\doibase 10.1029/2012GL051841} {\bibfield  {journal}
  {\bibinfo  {journal} {Geophys.\ Res.\ Lett.}\ }\textbf {\bibinfo {volume}
  {39}},\ \bibinfo {pages} {L11605} (\bibinfo {year} {2012})}\BibitemShut
  {NoStop}%
\bibitem [{\citenamefont {Espa}\ \emph {et~al.}(2014)\citenamefont {Espa},
  \citenamefont {Lacorata},\ and\ \citenamefont {{Di Nitto}}}]{ELDN:2014}%
  \BibitemOpen
  \bibfield  {author} {\bibinfo {author} {\bibfnamefont {S.}~\bibnamefont
  {Espa}}, \bibinfo {author} {\bibfnamefont {G.}~\bibnamefont {Lacorata}}, \
  and\ \bibinfo {author} {\bibfnamefont {G.}~\bibnamefont {{Di Nitto}}},\
  }\href@noop {} {\bibfield  {journal} {\bibinfo  {journal} {J. Phys.
  Oceanogr.}\ }\textbf {\bibinfo {volume} {44}},\ \bibinfo {pages} {632}
  (\bibinfo {year} {2014})}\BibitemShut {NoStop}%
\bibitem [{\citenamefont {Boffetta}\ \emph {et~al.}(2000)\citenamefont
  {Boffetta}, \citenamefont {Celani}, \citenamefont {Cencini}, \citenamefont
  {Lacorata},\ and\ \citenamefont {Vulpiani}}]{BCCLV:2000}%
  \BibitemOpen
  \bibfield  {author} {\bibinfo {author} {\bibfnamefont {G.}~\bibnamefont
  {Boffetta}}, \bibinfo {author} {\bibfnamefont {A.}~\bibnamefont {Celani}},
  \bibinfo {author} {\bibfnamefont {M.}~\bibnamefont {Cencini}}, \bibinfo
  {author} {\bibfnamefont {G.}~\bibnamefont {Lacorata}}, \ and\ \bibinfo
  {author} {\bibfnamefont {A.}~\bibnamefont {Vulpiani}},\ }\href@noop {}
  {\bibfield  {journal} {\bibinfo  {journal} {Chaos}\ }\textbf {\bibinfo
  {volume} {10}},\ \bibinfo {pages} {50} (\bibinfo {year} {2000})}\BibitemShut
  {NoStop}%
\bibitem [{\citenamefont {Cencini}\ and\ \citenamefont
  {Vulpiani}(2013)}]{CV:2013}%
  \BibitemOpen
  \bibfield  {author} {\bibinfo {author} {\bibfnamefont {M.}~\bibnamefont
  {Cencini}}\ and\ \bibinfo {author} {\bibfnamefont {A.}~\bibnamefont
  {Vulpiani}},\ }\href@noop {} {\bibfield  {journal} {\bibinfo  {journal}
  {Journal of Physics A}\ }\textbf {\bibinfo {volume} {46}},\ \bibinfo {pages}
  {(254019) 1} (\bibinfo {year} {2013})}\BibitemShut {NoStop}%
\bibitem [{\citenamefont {Taylor}(1921)}]{Taylor:1921}%
  \BibitemOpen
  \bibfield  {author} {\bibinfo {author} {\bibfnamefont {G.~I.}\ \bibnamefont
  {Taylor}},\ }\href@noop {} {\bibfield  {journal} {\bibinfo  {journal}
  {Proceedings of the London Mathematical Society}\ }\textbf {\bibinfo {volume}
  {20}},\ \bibinfo {pages} {196} (\bibinfo {year} {1921})}\BibitemShut
  {NoStop}%
\bibitem [{\citenamefont {Solomon}\ and\ \citenamefont
  {Gollub}(1988)}]{SG:1988}%
  \BibitemOpen
  \bibfield  {author} {\bibinfo {author} {\bibfnamefont {T.~H.}\ \bibnamefont
  {Solomon}}\ and\ \bibinfo {author} {\bibfnamefont {J.~P.}\ \bibnamefont
  {Gollub}},\ }\href@noop {} {\bibfield  {journal} {\bibinfo  {journal} {Phys.
  Rev. A}\ }\textbf {\bibinfo {volume} {38}},\ \bibinfo {pages} {6280}
  (\bibinfo {year} {1988})}\BibitemShut {NoStop}%
\bibitem [{\citenamefont {Chirikov}(1979)}]{Chirikov:1979}%
  \BibitemOpen
  \bibfield  {author} {\bibinfo {author} {\bibfnamefont {B.~V.}\ \bibnamefont
  {Chirikov}},\ }\href@noop {} {\bibfield  {journal} {\bibinfo  {journal}
  {Phys. Rep.}\ }\textbf {\bibinfo {volume} {52}},\ \bibinfo {pages} {263}
  (\bibinfo {year} {1979})}\BibitemShut {NoStop}%
\bibitem [{\citenamefont {Boffetta}\ \emph {et~al.}(1999)\citenamefont
  {Boffetta}, \citenamefont {Celani}, \citenamefont {Crisanti},\ and\
  \citenamefont {Vulpiani}}]{BCCV:1999}%
  \BibitemOpen
  \bibfield  {author} {\bibinfo {author} {\bibfnamefont {G.}~\bibnamefont
  {Boffetta}}, \bibinfo {author} {\bibfnamefont {A.}~\bibnamefont {Celani}},
  \bibinfo {author} {\bibfnamefont {A.}~\bibnamefont {Crisanti}}, \ and\
  \bibinfo {author} {\bibfnamefont {A.}~\bibnamefont {Vulpiani}},\ }\href@noop
  {} {\bibfield  {journal} {\bibinfo  {journal} {Phys. Rev. E}\ }\textbf
  {\bibinfo {volume} {60}},\ \bibinfo {pages} {6734} (\bibinfo {year}
  {1999})}\BibitemShut {NoStop}%
\end{thebibliography}
\end{document}